# 3D-DXA Cortical and Trabecular Parameters: Agreement Between Hologic Densitometers in Clinical Practice

Marta I. Bracco[1], Jorge Malouf[2], Laurent Maimoun[3,4], Xavier Nogues[5], Jean Paul Roux[6], François DuBoeuf[6], Ludovic Humbert[1]

[1]3D-Shaper Medical, Barcelona, Spain; [2]Mineral Metabolism Unit, Grup Creu Groga, Mataró, Spain; [3]Service de Médecine Nucléaire, Hôpital Lapeyronie, CHU Montpellier, France; [4]Physiologie et Médecine Expérimentale du Cœur et des Muscles (PhyMedEx), INSERM, CNRS, Université de Montpellier (UM) ; [5]Hospital del Mar Research Institute ,CIBERFES, Pompeu Fabra University  Barcelona, Spain; [6]INSERM UMR 1033, Université Claude Bernard-Lyon 1, Lyon, France;

**Abstract**

*Background* Three-dimensional dual-energy X-ray absorptiometry reconstructs three-dimensional maps of the proximal femur's density distribution from standard hip scans, enabling the estimation of trabecular and cortical bone parameters. The aim of this study was to assess the agreement of these three-dimensional cortical and trabecular femur parameters across different series and models of Hologic densitometers.

*Methodology* The study cohort was composed of 103 women and men recruited from four clinical centers in Spain and France. Subjects had duplicated hip scans using different Hologic scanners from the Horizon, Discovery, and QDR4500 series. Analyses were performed using 3D-Shaper software. Inter-scanner agreement was evaluated using Deming regression and Bland-Altman analysis.

*Results* The parameters demonstrated strong inter-device agreement across all clinical centers and scanner models, with coefficients of determination greater than 0.91. Absolute biases were less than 2.5 mg/cm$^3$ for integral volumetric bone mineral density, less than 2.9 mg/cm$^3$ for trabecular volumetric bone mineral density, and less than 1.7 mg/cm$^2$ for cortical surface bone mineral density. No statistically significant bias was found between parameters obtained from different scanners. Furthermore, the observed bias was lower than the expected least significant change, indicating that inter-scanner variability across these devices is not clinically significant.

*Conclusions* This study demonstrated excellent agreement for standard and three-dimensional derived bone parameters at the hip across Hologic densitometers. These findings support their suitability for clinical use.

# Introduction

Dual-energy X-ray absorptiometry (DXA) is the current gold standard imaging technique for the clinical diagnosis of osteoporosis. DXA enables bone mineral density (BMD) estimation from a two-dimensional (2D) projection of bone, with minimal ionizing radiation exposure, and excellent measurement precision. Due to its 2D nature, standard DXA cannot capture volumetric density or differentiate between the cortical and trabecular bone compartments. While advanced imaging modalities such as quantitative computed tomography (QCT) and high-resolution peripheral QCT (HR-pQCT) can directly measure true volumetric density and separate these bone compartments, their routine clinical implementation is limited by high operational costs, lack of widespread availability, and significantly higher ionizing radiation doses [1], [2], [3]. To overcome this limitation, DXA-based three-dimensional modelling methods have been proposed, allowing clinicians access estimations of 3D compartment-specific bone parameters [4], [5], [6]. Currently, the only available clinical tool implementing this 3D-DXA technology is 3D-Shaper® software, which provides a subject-specific model of the proximal femur obtained by registering a 3D statistical model onto a standard hip DXA scan[4]. The software calculates estimates of integral volumetric BMD, trabecular volumetric BMD and cortical surface BMD. The intra-scanner precision of these parameters has been assessed in two previous studies [7], [8]. In addition, 3D-Shaper software has been evaluated in various conditions, such as obesity[9], anorexia nervosa [10], spinal cord injury [11], [12], and post-bariatric surgery [13], [14].

GE Healthcare (Madison, WI, USA) and Hologic (Marlborough, Massachusetts, USA) are the main manufacturers of DXA machines. Measurement agreement between DXA systems has been previously evaluated for standard DXA areal BMD (aBMD) [15], [16]. Assessing the agreement of 3D-DXA measurements when subjects are scanned using different DXA scanners is equally important. In a recent study, excellent agreement in 3D-DXA measurements was reported between Prodigy and iDXA densitometers from GE Healthcare [8]. However, agreement between DXA scanners from Hologic has not yet been investigated. Differences in technological characteristics across Hologic scanner generations could potentially influence 3D-DXA results. Therefore, the aim of the present study is to evaluate the inter-scanner variability of 3D-DXA integral, cortical and trabecular bone parameters across Hologic DXA scanners.

## Material and methods

*Subjects and DXA Measurements*

Subjects from four clinical sites underwent duplicated hip scans using different Hologic DXA scanners. The clinical sites participating in the study were:

- Hospital de la Santa Creu i Sant Pau, Barcelona, Spain, referred to as *Hospital Sant Pau*;
- Hôpital Lapeyronie, CHRU Montpellier et Université Montpellier, Montpellier, France;
- Hospices Civils de Lyon, Hôpital Edouard Herriot, Service de Rhumatologie et Pathologie Osseuse, Lyon, France, referred to as *Hôpital Herriot*;
- Hospital del Mar, Barcelona, Spain.

Across all centers, subjects aged 20 years and older referred for routine bone health assessments were recruited. In accordance with standard DXA protocols, individuals were excluded if they presented with surgical implants.

Participants were scanned using Hologic Discovery W and Horizon A at *Hospital Sant Pau*, QDR 4500A and Horizon A at *Hôpital Lapeyronie*, Discovery A and Horizon A at *Hôpital Herriot*, and QDR 4500 SL and Horizon Wi at *Hospital del Mar*. Total hip areal BMD (aBMD) was calculated using the APEX® Software (Hologic, Marlborough, Massachusetts). Device calibration was performed daily according to the manufacturer's instructions.

*3D-DXA Analysis*

3D-DXA analyses were performed using 3D-Shaper® software (v2.14, 3D-Shaper Medical, Barcelona, Spain). 3D-DXA calculates a 3D model of the femoral shape and density distribution based on statistical modelling and image registration, as previously described [4]. Three measurements characterize the 3D femur at the total hip region: integral volumetric BMD (integral vBMD, expressed in $mg/cm^3$), cortical surface BMD (sBMD, expressed in $mg/cm^2$), and trabecular volumetric BMD (trabecular vBMD, expressed in $mg/cm^3$).

*Statistical Analysis*

Standard DXA and 3D-Shaper parameters assessed by each pair of scanners were compared using Deming regression and Bland and Altman analysis. From the Bland-Altman analysis, the interval of agreement at 95% (Limits of Agreement, LoA) and the bias were extracted. The LoA was reported as half of the width of the interval (i.e. the distance from the mean difference). Bias was reported as both the signed mean difference and the relative bias, expressed as a percentage of the mean value from scanner #1. The root mean squared error (RMSE) was also calculated for each parameter. Paired t-test was used to determine if the mean difference (bias) between measurements obtained using each pair of DXA scanners was significantly different from zero.

## Results

*Subject Characteristics and DXA analysis.* Table 1 presents the subjects characteristics and DXA scanners used in each clinical site.

*Table 1. Subject characteristics and DXA scanners used in the clinical sites participating in the study. Quantitative variables are reported as mean (standard deviation) [range]*

|  | Hospital Sant Pau | Hôpital Lapeyronie | Hôpital Herriot | Hospital del Mar |
|---|---|---|---|---|
| N | 35 | 35 | 15 | 18 |
| Sex | 30 women, 5 men | 28 women, 7 men | 11 women, 4 men | 16 women, 2 men |
| Age | 51.3 (12.6) [28.2-80.2] | 48.3 (10.9) [23.6-60.7] | 45.7 (11.9) [25.7-64.8] | 62.3 (17.3) [21.4-89.8] |
| Race | 35 White | 34 White<br>1 Black | 13 White<br>2 Black | 18 White |
| Height (cm) | 163.7 (9.2) [148.0-183.0] | 166.1 (7.1) [155.0-192.0] | 167.6 (8.9) [154.0-182.0] | 158.8 (7.5) [145.0-174.0] |
| Weight (kg) | 68.2 (13.7) [50.0-98.0] | 63.1 (10.7) [46.0-93.0] | 70.0 (16.2) [49.0-98.0] | 63.9 (13.0) [42.0-87.0] |
| BMI | 25.4 (4.1) [19.3-38.3] | 22.8 (3.0) [18.2-34.6] | 24.7 (4.5) [18.7-31.5] | 25.4 (5.3) [17.1-36.2] |
| Femur side | Right | Left | Right | Left |
| Scanner #1 | Hologic Discovery W | Hologic QDR 4500A | Hologic Discovery A | Hologic QDR 4500SL |
| APEX version | 3.3.0.1 | N/A | 3.3.0.1 | 2.4.2 |
| Scan mode | *Fast Array* | *Fast Array* | *Array* | *Array* |
| Scanner #2 | Hologic Horizon A | Hologic Horizon A | Hologic Horizon A | Hologic Horizon Wi |
| APEX version | 5.6.1.3 | 5.6.0.3 | 5.6.0.7 | 5.6.0.5 |
| Scan mode | *Fast Array* | *Fast Array* | *Array* | *Fast Array* |
| aBMD (g/cm$^2$) [1] | 0.900 (0.144) | 0.876 (0.146) | 0.998 (0.199) | 0.804 (0.147) |
| Integral vBMD (mg/cm$^3$) [1] | 313.4 (57.0) | 329.1 (62.7) | 357.6 (67.2) | 273.9 (54.3) |
| Trabecular vBMD (mg/cm$^3$) [1] | 181.2 (41.2) | 174.8 (47.8) | 211.9 (47.1) | 149.6 (36.4) |
| Cortical sBMD (mg/cm$^2$) [1] | 151.7 (25.6) | 161.9 (28.3) | 174.4 (31.4) | 140.4 (25.1) |

[1]*Measured with scanner #1;*

*Inter-scanner agreement for areal BMD.* aBMD agreements between the Hologic scanner pairs were strongly correlated, with $R^2$ values ranging from 0.97 to 1.00 depending on the clinical site, as reported in Table 2. Deming regression and Bland-Altman plots for aBMD are reported for each clinical center in *Figure* 1*, Figure* 2*, Figure* 3*,* and *Figure* 4. (top). The mean difference (bias) between aBMD values assessed with the two scanners was below 0.012 g/cm$^2$, corresponding to 1.5% of the mean value, and was not statistically significant except for the cohort from *Hospital San Pau* clinical site ($p < 0.01$). The Limits of Agreement (LoA) for aBMD measurements ranged from 0.013 g/cm$^2$ to 0.031 g/cm$^2$. The slope of the regression line was between 0.92 and 1.0, and the intercept ranged between 0.004 g/cm$^2$ and 0.055 g/cm$^2$

*Inter-scanner agreement for 3D-DXA.* Strong agreements were also obtained for all 3D-DXA parameters across the different Hologic scanner combinations, as shown in Table 2, with $R^2$ above 0.91 and regression line slopes between 0.92 and 1.03 for all parameters. Biases had absolute values < 2.5 mg/cm$^3$ for integral vBMD, < 2.9 mg/cm$^3$ for trabecular vBMD, and < 1.7 mg/cm$^2$ for cortical sBMD, corresponding to 0.7 %, 1.6 % and 1.0 % of the mean value respectively, and they were not statistically significant. Deming regression and Bland-Altman plots for integral vBMD, trabecular vBMD and cortical sBMD are reported for each clinical center in *Figure* 1*, Figure* 2*, Figure* 3*,* and *Figure* 4.

*Table 2*. Inter-scanner agreements for standard DXA and 3D-DXA parameters.

|  |  | Hospital Sant Pau | Hôpital Lapeyronie | Hôpital Herriot | Hospital del Mar |
|---|---|---|---|---|---|
| **aBMD** | Bias: |  |  |  |  |
|  | value (g/cm$^2$) | - 0.011 | 0.001 | - 0.007 | - 0.012 |
|  | relative (%) | - 1.2 | 0.1 | - 0.7 | - 1.5 |
|  | p-value | 0.006 | 0.665 | 0.102 | 0.106 |
|  | Slope (-) | 0.95 | 1.0 | 0.96 | 0.92 |
|  | Intercept (g/cm²) | 0.031 | 0.004 | 0.030 | 0.055 |
|  | RMSE (g/cm²) | 0.023 | 0.011 | 0.016 | 0.030 |
|  | LoA (g/cm²) | 0.028 | 0.031 | 0.013 | 0.022 |
|  | R² | 0.98 | 0.99 | 1.00 | 0.97 |
| **Integral vBMD** | Bias: |  |  |  |  |
|  | value (mg/cm³) | - 2.5 | - 0.5 | - 0.1 | 0.7 |
|  | relative (%) | - 0.7 | - 0.1 | - 0.0 | 0.2 |
|  | p-value | 0.089 | 0.742 | 0.923 | 0.767 |
|  | Slope (-) | 0.99 | 0.99 | 1.0 | 0.96 |
|  | Intercept (mg/cm³) | - 0.8 | 1.5 | 0.5 | 12.6 |
|  | RMSE (mg/cm³) | 8.8 | 9.0 | 4.8 | 10.1 |
|  | LoA (mg/cm³) | 14.8 | 19.4 | 9.7 | 18.6 |
|  | R² (-) | 0.98 | 0.98 | 0.99 | 0.96 |
| **Trabecular vBMD** | Bias: |  |  |  |  |
|  | value (mg/cm³) | - 2.9 | 0.1 | - 1.0 | 0.9 |
|  | relative (%) | - 1.6 | 0.0 | - 0.5 | 0.6 |
|  | p-value | 0.080 | 0.948 | 0.493 | 0.718 |
|  | Slope (-) | 0.95 | 0.92 | 1.0 | 0.96 |
|  | Intercept (mg/cm³) | 5.4 | 13.4 | - 1.0 | 6.9 |
|  | RMSE (mg/cm³) | 9.8 | 7.9 | 5.7 | 10.6 |
|  | LoA (mg/cm³) | 21.4 | 17.4 | 10.7 | 18.8 |
|  | R² (-) | 0.95 | 0.98 | 0.99 | 0.91 |
| **Cortical sBMD** | Bias: |  |  |  |  |
|  | value (mg/cm²) | - 0.8 | - 1.7 | 1.4 | - 0.8 |
|  | relative (%) | - 0.5 | - 1.0 | 0.8 | - 0.6 |
|  | p-value | 0.250 | 0.090 | 0.109 | 0.620 |
|  | Slope (-) | 0.99 | 1.04 | 1.02 | 1.03 |
|  | Intercept (mg/cm²) | - 0.1 | - 7.5 | - 2.4 | - 4.2 |
|  | RMSE (mg/cm²) | 4.1 | 6.0 | 3.5 | 6.6 |
|  | LoA (mg/cm²) | 7.8 | 12.9 | 7.3 | 12.0 |
|  | R² (-) | 0.97 | 0.96 | 0.99 | 0.93 |

[1]*Measured with scanner #1; $R^2$, coefficient of determination; LoA, limits of agreement, RMSE, root mean square error*

## Discussion

This is the first study to broadly evaluate the inter-scanner variability for 3D-DXA parameters across multiple Hologic densitometers.

Pairwise correlations between different scanner models were found to be very strong, with values of $R^2$ above 0.9 for all parameters. For aBMD, bias did not exceed 0.012 g/cm$^2$, while LoA peaked at 0.031 g/cm$^2$. With the exception of *Hospital Sant Pau*, the observed bias did not reach statistical significance at any of the clinical sites. A previous study analyzed the inter-scanner variability of aBMD between Hologic QDR4500W and Discovery Wi DXA scanner models, and their results were in line with the current study, with a bias of 0.007 g/cm$^2$ and LoA 0.038 g/cm$^2$ for total hip aBMD [16]. In that study, the authors also reported comparable inter-scanner and intra-scanner variability, concluding that the bias was clinically insignificant. Another previous study looked at the cross-correlation between Hologic Discovery A and Horizon A scanner models, finding a very good agreement between BMD measurements, with $R^2 > 0.98$ and a bias around -0.01 g/cm$^2$ at the total hip [17]. These values are aligned with our results for the same scanner models, with an $R^2$ of 1.0 and a bias of -0.007 g/cm$^2$. [15]

Regarding 3D-DXA parameters, when analyzing the inter-scanner variations, the biases observed across all clinical sites were below 2.9 mg/cm$^3$ for both integral and trabecular vBMD, and below 1.7 mg/cm$^2$ for cortical sBMD. These values are in line with those found for GE Healthcare devices in a previous study, where bias was below 3.0 mg/cm$^3$ for integral and trabecular vBMD and below 1.0 mg/cm$^2$ for cortical sBMD [8]. No statistically significant bias was observed for any 3D-DXA parameter across all clinical sites.

To assess whether the variability we observed in standard DXA and 3D-DXA parameters was clinically significant, the biases found in this study can be compared to the measurement precision. The Least Significant Change (LSC) was previously established for repeated measurements on the Hologic Discovery W scanner included in this study (*San Pau* clinical center), resulting in values of 0.03 g/cm² for total femur aBMD, 10.4 mg/cm³ for integral vBMD, 9.6 mg/cm³ for trabecular vBMD, and 6.3 mg/cm² for cortical sBMD [7]. Because the observed inter-scanner biases fall well below their corresponding LSC, for both standard DXA and 3D-DXA parameters, the inter-scanner biases can be considered as clinically irrelevant. Consequently, value adjustment is not needed when assessing 3D-DXA parameters across different Hologic scanners.

Since 3D-DXA analyses are derived from DXA scans, it is also important to consider how the underlying inter-scanner aBMD bias might influence the bias observed in 3D-DXA parameters. Notably, the aBMD bias observed at *Hospital Sant Pau* and *Hospital Del Mar* was substantially higher compared to *Hôpital Lapeyronie*. However, this greater difference in 2D measurements did not translate into a correspondingly higher bias in the 3D-DXA parameters for those sites, suggesting that the inter-scanner error is attenuated by the 3D-DXA analysis. The superior consistency of 3D-DXA over standard DXA measurements could be due to its ability to automatically correct for changes in patient positioning. Similar findings were reported in the inter-scanner and precision study using GE Healthcare scanners [8].

Our study has a few limitations that must be acknowledged. First, the sample size evaluated across the different clinical sites was relatively small, especially at *Hôpital Herriot* and *Hospital Del Mar* clinical sites, which may limit the statistical power and broader generalizability of our findings. Secondly, within our analysis, we did not test the precision of each individual device included in the analysis. Consequently, to determine clinical significance of inter-scanner biases, we had to rely on precision metrics from a single Hologic model, rather than device-specific baselines. In addition, we did not consider all the possible scanner models and combinations due to the lack of data. Nevertheless, because the devices evaluated in our study encompass the main Hologic scanners series (Horizon, Discovery, and QDR4500) and models (A, W, SL, and Wi), these findings are expected to generalize across all Hologic devices.

In conclusion, 3D-DXA-derived bone parameters were found to have low inter-scanner variability across different Hologic models, with no statistically or clinically significant biases. Although the use of multiple Hologic scanner models is generally not recommended for longitudinal patient monitoring, the findings of this study support that, at the cohort level, 3D-DXA parameters can be applied across different Hologic models without the need for value adjustment.

## References

[1]   C. Shuhart, A. Cheung, R. Gill, L. Gani, H. Goel, and A. Szalat, "Executive Summary of the 2023 Adult Position Development Conference of the International Society for Clinical Densitometry: DXA Reporting, Follow-up BMD Testing and Trabecular Bone Score Application and Reporting," *Journal of Clinical Densitometry*, vol. 27, no. 1, p. 101435, Jan. 2024, doi: 10.1016/j.jocd.2023.101435.

[2]   K. Engelke *et al.*, "Clinical Use of Quantitative Computed Tomography (QCT) of the Hip in the Management of Osteoporosis in Adults: The 2015 ISCD Official Positions-Part I," *Journal of Clinical Densitometry*, vol. 18, no. 3, pp. 338–358, 2015, doi: 10.1016/j.jocd.2015.06.012.

[3]   S. Gazzotti *et al.*, "High-resolution peripheral quantitative computed tomography:

research or clinical practice?," 2023, *British Institute of Radiology*. doi: 10.1259/bjr.20221016.

[4] L. Humbert *et al.*, "3D-DXA: Assessing the Femoral Shape, the Trabecular Macrostructure and the Cortex in 3D from DXA images," *IEEE Trans. Med. Imaging*, vol. 36, no. 1, pp. 27–39, Jan. 2017, doi: 10.1109/TMI.2016.2593346.

[5] S. P. Väänänen, L. Grassi, G. Flivik, J. S. Jurvelin, and H. Isaksson, "Generation of 3D shape, density, cortical thickness and finite element mesh of proximal femur from a DXA image," *Med. Image Anal.*, vol. 24, no. 1, pp. 125–134, Aug. 2015, doi: 10.1016/j.media.2015.06.001.

[6] L. Grassi, S. P. Väänänen, M. Ristinmaa, J. S. Jurvelin, and H. Isaksson, "Prediction of femoral strength using 3D finite element models reconstructed from DXA images: validation against experiments," *Biomech. Model. Mechanobiol.*, vol. 16, no. 3, pp. 989–1000, Jun. 2017, doi: 10.1007/s10237-016-0866-2.

[7] L. Humbert *et al.*, "3D Analysis of Cortical and Trabecular Bone From Hip DXA:Precision and Trend Assessment Interval in Postmenopausal Women," *Journal of Clinical Densitometry*, vol. 22, no. 2, pp. 214–218, Apr. 2019, doi: 10.1016/j.jocd.2018.05.001.

[8] D. Krueger *et al.*, "3D-DXA Cortical and Trabecular Parameters; Agreement and Precision Between GE Healthcare Prodigy and iDXA Densitometers," Mar. 04, 2026. doi: 10.64898/2026.03.04.26347524.

[9] L. Maïmoun *et al.*, "Modification of bone mineral density, bone geometry and volumetric BMD in young women with obesity," *Bone*, vol. 150, p. 116005, Sep. 2021, doi: 10.1016/j.bone.2021.116005.

[10]  L. Maïmoun *et al.*, "Effect of Anorexia Nervosa on Volumetric Bone Mineral Density.," *Calcif. Tissue Int.*, vol. 116, no. 1, p. 132, Oct. 2025, doi: 10.1007/s00223-025-01442-1.

[11]  L. Gifre *et al.*, "Analysis of the evolution of cortical and trabecular bone compartments in the proximal femur after spinal cord injury by 3D-DXA," *Osteoporosis International*, vol. 29, no. 1, pp. 201–209, Jan. 2018, doi: 10.1007/s00198-017-4268-9.

[12]  L. Maïmoun *et al.*, "Alteration of Volumetric Bone Mineral Density Parameters in Men with Spinal Cord Injury," *Calcif. Tissue Int.*, vol. 113, no. 3, pp. 304–316, Jun. 2023, doi: 10.1007/s00223-023-01110-2.

[13]  L. Maïmoun *et al.*, "Modification of bone architecture following sleeve gastrectomy: a five-year follow-up," *Journal of Bone and Mineral Research*, vol. 40, no. 2, pp. 251–261, Feb. 2025, doi: 10.1093/jbmr/zjae202.

[14]  C. Gómez-Vaquero *et al.*, "Bone loss after bariatric surgery is observed mainly in the hip trabecular compartment and after hypoabsorptive techniques," *Bone*, vol. 190, Jan. 2025, doi: 10.1016/j.bone.2024.117270.

[15]  D. Krueger, N. Vallarta-Ast, M. Checovich, D. Gemar, and N. Binkley, "BMD Measurement and Precision: A Comparison of GE Lunar Prodigy and iDXA Densitometers," *Journal of Clinical Densitometry*, vol. 15, no. 1, pp. 21–25, 2012, doi: 10.1016/j.jocd.2011.08.003.

[16]  M. K. Covey, D. L. Smith, J. K. Berry, and E. D. Hacker, "Importance of cross-calibration when replacing DXA Scanners: QDR4500W and Discovery Wi," *J. Nurs. Meas.*, vol. 16, no. 3, pp. 155–170, 2008, doi: 10.1891/1061-3749.16.3.155.

[17]  S. Promma, "Cross calibration between two dual energy X-ray absorptiometry systems: Horizon A and Discovery A," *Chulalongkorn Medical Journal*, vol. 66, no. 2, pp. 145–151,

*Figure 1*. Deming regression (left) & Bland-Altman plot (right) between Hologic Discovery W and Horizon A devices at Hospital Sant Pau (Barcelona, Spain) for aBMD and 3D-DXA parameters. In Bland-Altman plot, the blue line represents the bias while the dotted blue lines represent the LOA interval at 95%.

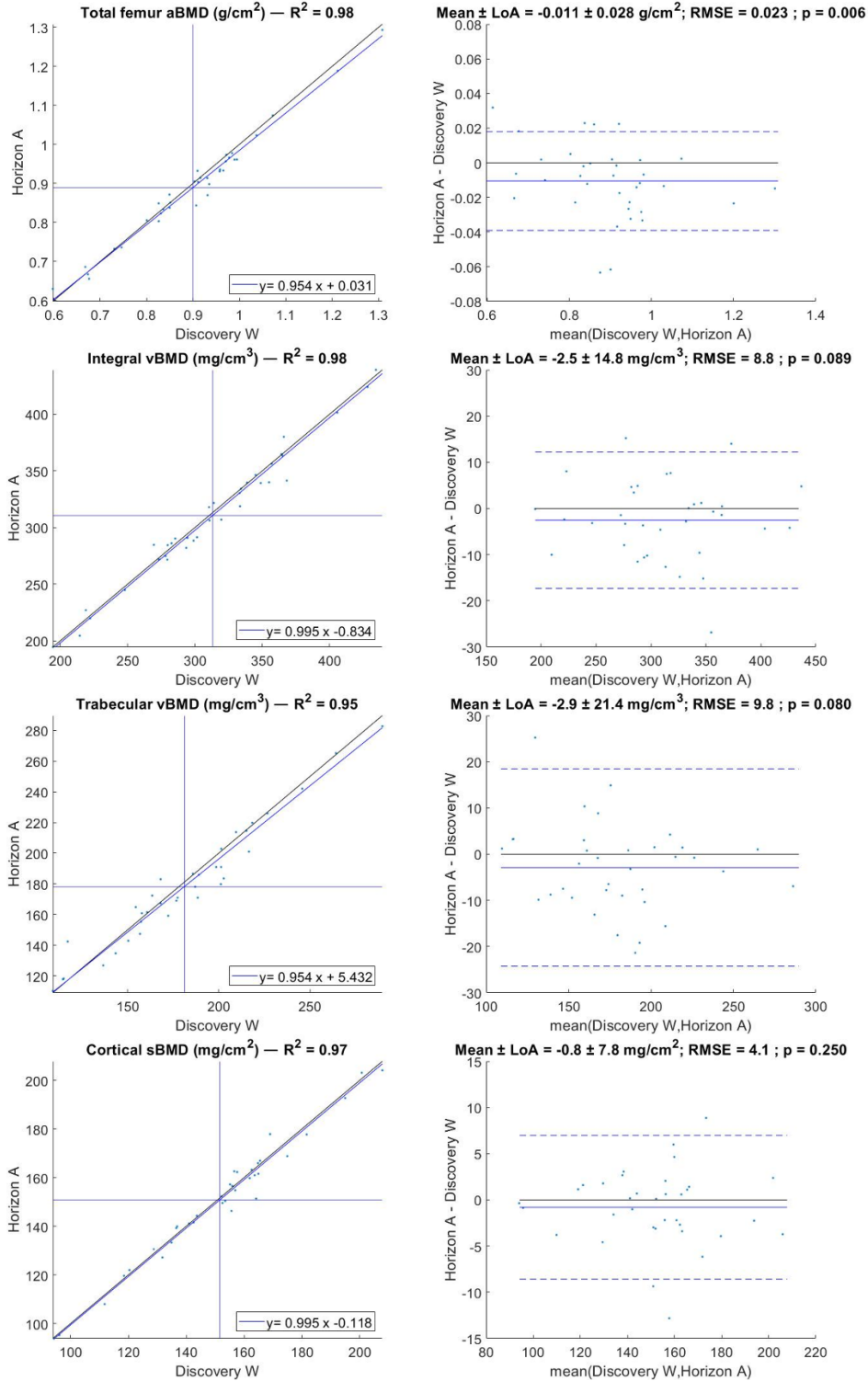

*aBMD, areal Bone Mineral Density; R², measurement of the goodness of fit; Mean, bias between the two devices; LoA, Limit of Agreement; RMSE, Root Mean Square Error; p-value, probability of paired t-test for mean difference=0*

*Figure 2.* Deming regression (left) & Bland-Altman plot (right) between Hologic QDR 4500A and Horizon A devices at Hôpital Lapeyronie (Montpellier, France), for aBMD and 3D-DXA parameters. In Bland-Altman plot, the blue line represents the bias while the dotted blue lines represent the LOA interval at 95%.

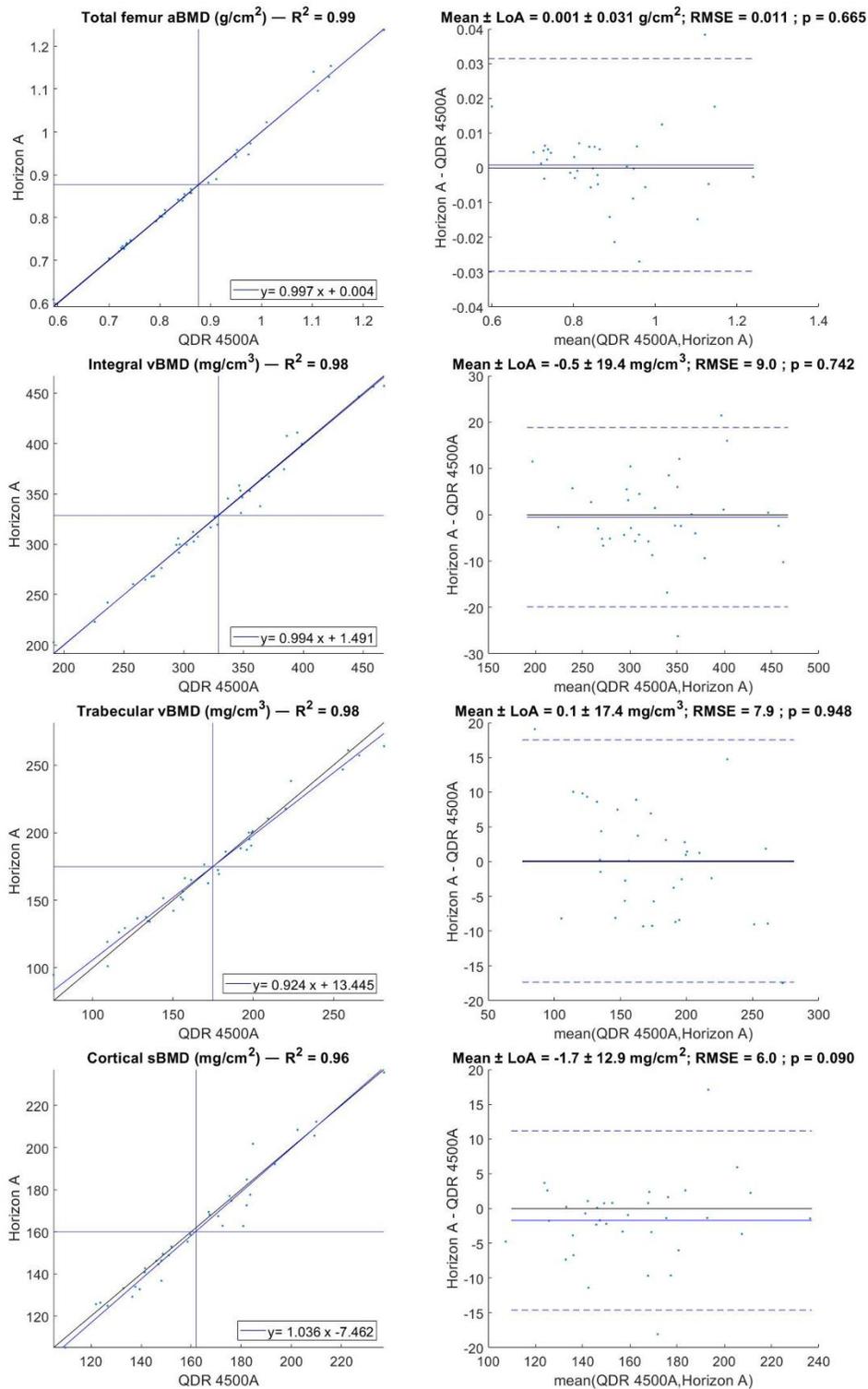

*aBMD, areal Bone Mineral Density; R², measurement of the goodness of fit; Mean, bias between the two devices; LoA, Limit of Agreement; RMSE, Root Mean Square Error; p-value, probability of paired t-test for mean difference=0*

*Figure 3.* Deming regression (left) & Bland-Altman plot (right) between Hologic Discovery A and Horizon A devices at *Hôpital Herriot* for aBMD and 3D-DXA parameters. In Bland-Altman plot, the blue line represents the bias while the dotted blue lines represent the LOA interval at 95%.

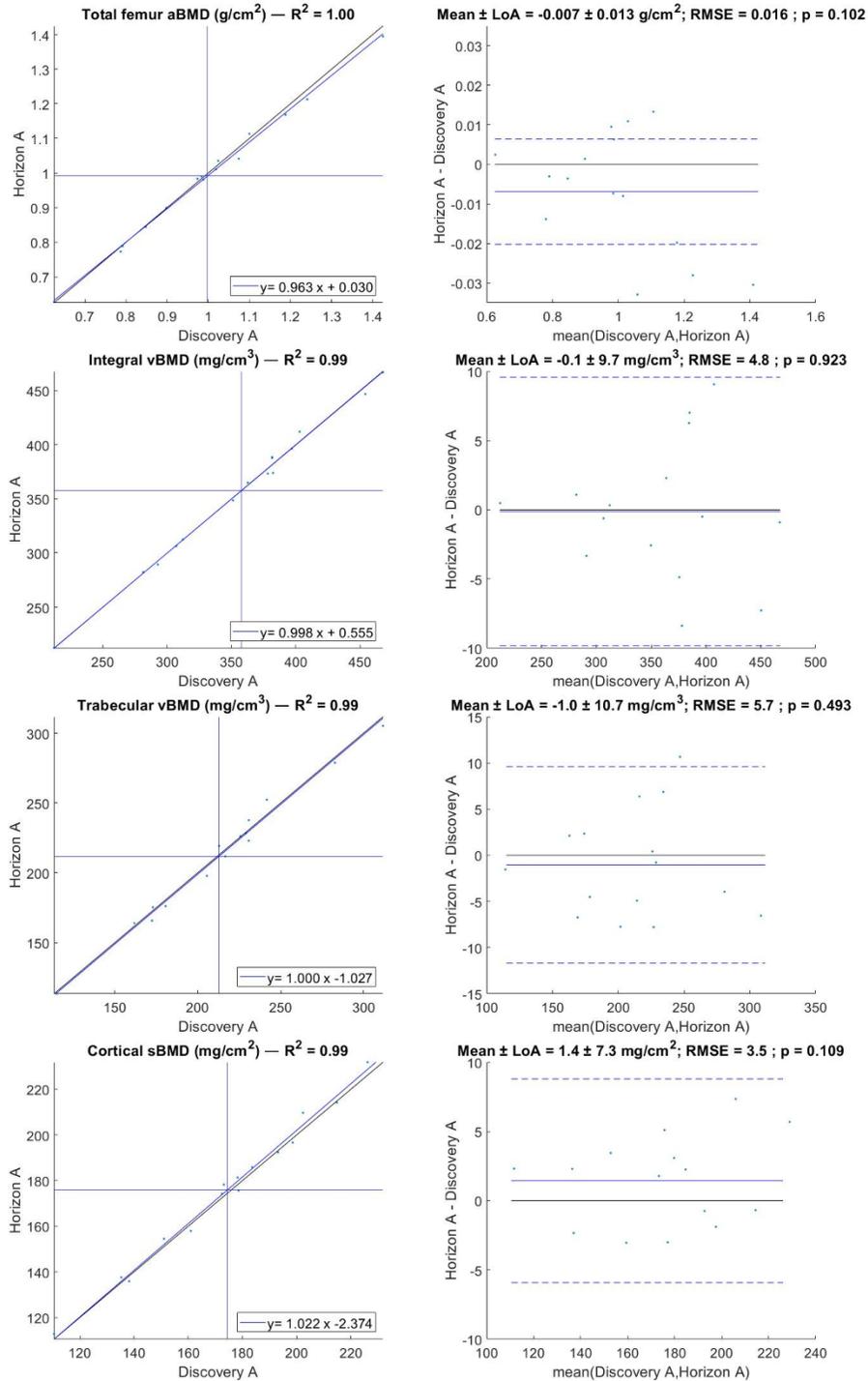

*aBMD, areal Bone Mineral Density; R², measurement of the goodness of fit; Mean, bias between the two devices; LoA, Limit of Agreement; RMSE, Root Mean Square Error; p-value, probability of paired t-test for mean difference=0*

*Figure 4.* Deming regression (left) & Bland-Altman plot (right) between Hologic QDR 4500SL and Horizon Wi devices at Hospital Del Mar (Barcelona, Spain) for aBMD and 3D-Shaper parameters. In Bland-Altman plot, the blue line represents the bias while the dotted blue lines represent the LOA interval at 95%.

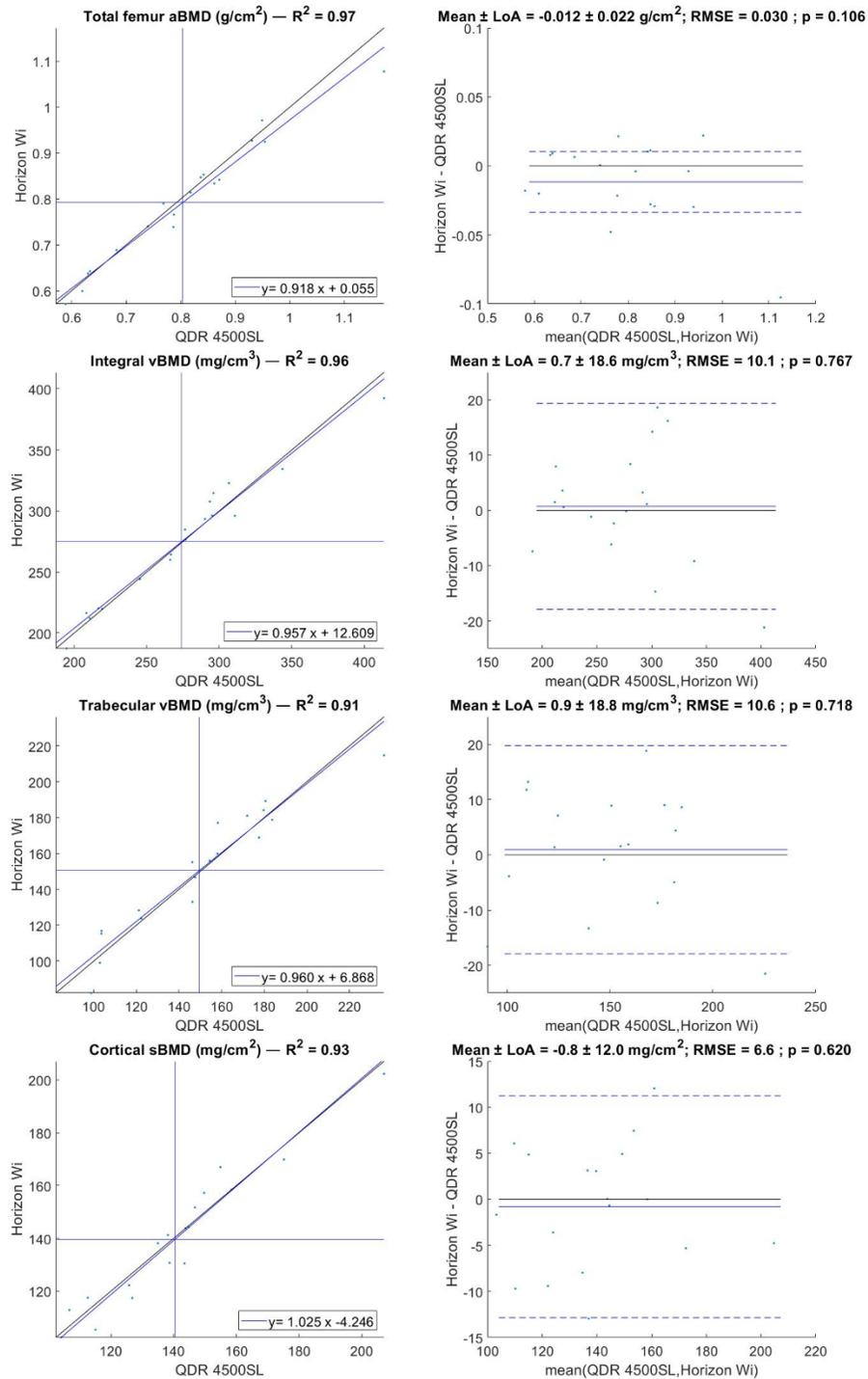

aBMD, areal Bone Mineral Density; $R^2$, measurement of the goodness of fit; Mean, bias between the two devices; LoA, Limit of Agreement; RMSE, Root Mean Square Error; p-value, probability of paired t-test for mean difference=0